\documentclass[twocolumn]{aastex631}

\begin{document}

\title{SN\,2023ixf in Messier 101: the twilight years of the progenitor as seen by Pan-STARRS}

\author[0000-0003-4175-4960]{Conor L. Ransome}
\affiliation{Center for Astrophysics \textbar{} Harvard \& Smithsonian, 60 Garden Street, Cambridge, MA 02138-1516, USA}

\author[0000-0002-1125-9187]{V.~Ashley~Villar}
\affiliation{Center for Astrophysics \textbar{} Harvard \& Smithsonian, 60 Garden Street, Cambridge, MA 02138-1516, USA}

\author[0000-0001-9229-8833]{Anna~Tartaglia}
\affiliation{Department of Astronomy and Astrophysics, The Pennsylvania State University, University Park, PA 16802, USA}

\author[0009-0002-8655-2609]{Sebastian~Javier~Gonzalez}
\affiliation{Department of Astronomy, University of California, Berkeley, CA 94720, USA}

\author[0000-0002-3934-2644]{Wynn~V.~Jacobson-Gal\'{a}n}
\affiliation{Department of Astronomy, University of California, Berkeley, CA 94720, USA}

\author[0000-0002-5740-7747]{Charles~D.~Kilpatrick}
\affiliation{Center for Interdisciplinary Exploration and Research in Astrophysics (CIERA) and Department of Physics and Astronomy, Northwestern University, Evanston, IL 60208, USA}

\author[0000-0003-4768-7586]{Raffaella~Margutti}
\affiliation{Department of Astronomy, University of California, Berkeley, CA 94720, USA}
\affiliation{Department of Physics, University of California, Berkeley, CA 94720, USA}

\author[0000-0002-2445-5275]{Ryan~J.~Foley}
\affiliation{Department of Astronomy and Astrophysics, University of California, Santa Cruz, CA 95064, USA}

\author[0000-0002-6741-983X]{Matthew~Grayling}
\affiliation{Institute of Astronomy and Kavli Institute for Cosmology, Madingley Road, Cambridge CB3 0HA, UK}

\author[0000-0003-3656-5268]{Yuan~Qi~Ni}
\affiliation{David A. Dunlap Department of Astronomy and Astrophysics, University of Toronto, 50 St. George Street, Toronto, Ontario, M5S 3H4, Canada}

\author[0000-0003-0381-1039]{Ricardo~Yarza}
\affiliation{Department of Astronomy and Astrophysics, University of California, Santa Cruz, CA 95064, USA}

\author{Christine~Ye}
\affiliation{Physics Department, Stanford University, Stanford, CA, USA}

\author[0000-0002-4449-9152]{Katie~Auchettl}
\affiliation{Department of Astronomy and Astrophysics, University of California, Santa Cruz, CA 95064, USA}
\affiliation{School of Physics, The University of Melbourne, VIC 3010, Australia}

\author[0000-0001-5486-2747]{Thomas~de~Boer}
\affiliation{Institute for Astronomy, University of Hawaii, 2680 Woodlawn Drive, Honolulu, HI 96822, USA}

\author[0000-0001-6965-7789]{Kenneth~C.~ Chambers}
\affiliation{Institute for Astronomy, University of Hawaii, 2680 Woodlawn Drive, Honolulu, HI 96822, USA}

\author[0000-0003-4263-2228]{David~A.~Coulter}
\affiliation{Space Telescope Science Institute, Baltimore, MD 21218, USA}

\author[0000-0001-7081-0082]{Maria~R.~Drout}
\affiliation{David A. Dunlap Department of Astronomy and Astrophysics, University of Toronto, 50 St. George Street, Toronto, Ontario, M5S 3H4, Canada}

\author[0000-0002-6886-269X]{Diego~Farias}
\affiliation{DARK, Niels Bohr Institute, University of Copenhagen, Jagtvej 128, 2200 Copenhagen, Denmark}

\author[0000-0002-8526-3963]{Christa~Gall}
\affiliation{DARK, Niels Bohr Institute, University of Copenhagen, Jagtvej 128, 2200 Copenhagen, Denmark}

\author[0000-0003-1015-5367]{Hua~Gao}
\affiliation{Institute for Astronomy, University of Hawaii, 2680 Woodlawn Drive, Honolulu, HI 96822, USA}

\author[0000-0003-1059-9603]{Mark~E.~Huber}
\affiliation{Institute for Astronomy, University of Hawaii, 2680 Woodlawn Drive, Honolulu, HI 96822, USA}

\author[0000-0003-2405-2967]{Adaeze~L.~Ibik}
\affiliation{David A. Dunlap Department of Astronomy and Astrophysics, University of Toronto, 50 St. George Street, Toronto, Ontario, M5S 3H4, Canada}

\author[0000-0002-6230-0151]{David~O.~Jones}
\affiliation{Gemini Observatory, NSF's NOIRLab, 670 N. A'ohoku Place, Hilo, Hawai'i, 96720, USA}

\author[0000-0003-2720-8904]{Nandita~Khetan}
\affiliation{DARK, Niels Bohr Institute, University of Copenhagen, Jagtvej 128, 2200 Copenhagen, Denmark}
\affiliation{School of Mathematics and Physics, The University of Queensland, QLD 4072, Australia}

\author[0000-0002-7272-5129]{Chien-Cheng~Lin}
\affiliation{Institute for Astronomy, University of Hawaii, 2680 Woodlawn Drive, Honolulu, HI 96822, USA}

\author[0000-0003-3727-9167]{Collin~A.~Politsch}
\affiliation{Institute of Astronomy and Kavli Institute for Cosmology, Madingley Road, Cambridge, CB3 0HA, UK}

\author[0000-0002-6248-398X]{Sandra~I.~Raimundo}
\affiliation{DARK, Niels Bohr Institute, University of Copenhagen, Jagtvej 128, 2200 Copenhagen, Denmark}
\affiliation{Department of Physics and Astronomy, University of Southampton, Highfield, Southampton SO17 1BJ, UK}

\author[0000-0002-4410-5387]{Armin~Rest}
\affiliation{Department of Physics and Astronomy, The Johns Hopkins University, Baltimore, MD 21218, USA}
\affiliation{Space Telescope Science Institute, Baltimore, MD 21218, USA}

\author[0000-0002-1341-0952]{Richard~J.~Wainscoat}
\affiliation{Institute for Astronomy, University of Hawaii, 2680 Woodlawn Drive, Honolulu, HI 96822, USA}

\author[0000-0002-0840-6940]{S.~Karthik~Yadavalli}
\affiliation{Center for Astrophysics \textbar{} Harvard \& Smithsonian, 60 Garden Street, Cambridge, MA 02138-1516, USA}

\author[0000-0002-0632-8897]{Yossef~Zenati}
\affiliation{Department of Physics and Astronomy, The Johns Hopkins University, Baltimore, MD 21218, USA}
\affiliation{Space Telescope Science Institute, Baltimore, MD 21218, USA}

\begin{abstract}

The nearby type II supernova, SN\,2023ixf in M\,101 exhibits signatures of early-time interaction with circumstellar material in the first week post-explosion. This material may be the consequence of prior mass loss suffered by the progenitor which possibly manifested in the form of a detectable pre-supernova outburst. We present an analysis of the long-baseline pre-explosion photometric data in $g$, $w$, $r$, $i$, $z$ and $y$ filters from Pan-STARRS as part of the Young Supernova Experiment, spanning $\sim$5,000 days. We find no significant detections in the Pan-STARRS pre-explosion light curve. We train a multilayer perceptron neural network to classify pre-supernova outbursts. We find no evidence of eruptive pre-supernova activity to a limiting absolute magnitude of $-7$. The limiting magnitudes from the full set of \textit{gwrizy} (average absolute magnitude $\approx\,$--8) data are consistent with previous pre-explosion studies. We use deep photometry from the literature to constrain the progenitor of SN\,2023ixf, finding that these data are consistent with a dusty red supergiant (RSG) progenitor with luminosity $\log\left(L/L_\odot\right)$\,$\approx$\,5.12 and temperature $\approx$\,3950\,K, corresponding to a mass of 14\,--\,20\,M$_\odot$.

\end{abstract}


\section{Introduction} \label{sec:intro}

Core collapse supernovae (CCSNe) are the explosive deaths of massive stars \citep[with $M_*\,\gtrsim\,8\,M_\odot$;][]{Woosley_2002}. Hydrogen-rich CCSNe, classified as type II SNe (SNe\,II) comprise $\sim$70\% of the observed CCSN population \citep[e.g.,][]{Li_2011, ysedr1, Tinyanont_2023}. SNe\,II make up the vast majority \citep[][]{VanDyk_2017} of pre-explosion progenitor detections via serendipitous imaging, e.g., SN\,2003gd \citep{Hendry_2005}, SN\,2013ej \citep{Fraser_IIP}, SN\,2017aew \citep{Kilpatrick_2018} and SN\,2022acko \citep{VanDyk_2023}. All of the observed progenitors of ``normal" SNe\,II (i.e. types IIP/L) have been red supergiants (RSGs) with masses that do not exceed $\sim$\,20\,M$_\odot$ \citep{Smartt_2009, Beasor_2020}. 

The remarkably proximate SN\,2023ixf ($\alpha$\,=\,14:03:38.56, $\delta$\,=\,+54:18:41.97, J2000) was discovered on 19 May 2023 by \citet{Itagaki_2023}. The host of SN\,2023ixf is M\,101 (also known as NGC\,5457 or the Pinwheel Galaxy), is at a distance of only 6.9\,Mpc (as measured via Cepheids; \citealt{Riess_2022}).  A classification spectrum from SPRAT on the Liverpool Telescope revealed SN\,2023ixf as a SN\,II \citep{Perley_2023ixf}. The discovery of SN\,2023ixf led to a sustained spectroscopic and photometric follow-up effort \citep[e.g.][]{Sgro_2022, Smith_2023_ixf, Bostroem_2023_ixf, Jacobson-Galan_2023_ixf, Berger_2023}. To date, these multi-wavelength follow-up observations and archival data examination have revealed detections of a dusty red supergiant (RSG) progenitor and signatures of interaction with circumstellar material \citep[e.g.][]{Pledger_2023, Jacobson-Galan_2023_ixf,Kilpatrick_2023_ixf,  VanDyk_2023, Smith_2023_ixf, Bostroem_2023_ixf, Niu_2023, Qin_2023, Xiang_2023,Koenig_2023, Soraisam_2023_ixf, Jencson_2023_ixf, Hiramatsu_2023, Berger_2023, Neustadt_2023_ixf, Vasylyev_2023_ixf, Prantik_2023, SinghTeja_2023, 2023arXiv230813101P, Grefenstette_2023_xray_23ixf,Zhang_2023, Guetta_2023_ixfneutrinos, Kong_2023, Yamanaka_2023, Li_2023}.

SNe\,II exhibiting interaction signatures, attributed to interaction with a confined, dense, slow and pre-existing circumstellar medium (CSM) are somewhat common. Around 30\% of SNe\,II show these flash-ionization features in addition to steep rises to peak, indicative of shock breakout out of dense CSM \citep{Bruch_2021, Forster_2018}. Similar early-time interaction is seen in SN\,2023ixf \citep{Jacobson-Galan_2023_ixf, Smith_2023_ixf, Bostroem_2023_ixf, Berger_2023, Grefenstette_2023_xray_23ixf, Chandra_2023, Mereminsky_2023, Kong_2023,2023arXiv230813101P}. The presence of flash ionization features in CCSNe suggests enhanced mass-loss rates in addition to supergiant winds in the final years of the life of their progenitors. While supergiant winds with a steady mass-loss rate $\dot{M}$\,$\sim$\,10$^{-6}$\,M$_\odot$\,yr$^{-1}$ are common in RSGs, these steady state mass-loss rates are too low to account for the mass stripping which leads to flash-ionization features \citep[e.g.,][]{Beasor_2020}. Furthermore, if supergiant winds are the primary mass-loss route for RSGs, one would expect an environmental metallicity dependence which is not seen for RSGs in M31 \citep[see,][]{McDonald_2022}. It is possible that enhanced mass loss modes such as ``superwinds" or outbursts driven by gravity waves with mass loss rates up to $\sim$\,10$^{-2}$\,M$_\odot$\,yr$^{-1}$ may help strip mass off a RSG progenitor \citep{Wu_2021,Davies_2022, Jacobson-Galan_2022}.

Whilst pre-SN mass-loss may be indirectly probed with followup spectroscopic observations \citep[e.g. via low velocity emission lines in spectra][]{Gal-Yam_2014}, outburst-like pre-SN activity may be directly observable. Models of pre-SN outbursts have predicted observable signatures lasting a few\,--\,to\,--\,hundreds of days with peak aboslute magnitudes M$_R$\,$\sim$\,--8.5 to --10 \citep{Davies_2022, Tsuna_2023}. While pre-SN mass loss is common in SNe\,IIn and ``regular" SNe\,II \citep[as inferred from light curve shapes, spectral features such as flash-ionization and X-ray observations, e.g.,][]{Ofek_2014,Forster_2018,Strotjohann_2020, Bruch_2021, 2023arXiv230813101P}, the luminous type II SN, SN\,2020tlf stands out as an example of a SN\,II which had a bright, detectable pre-explosion outburst. \citet{Jacobson-Galan_2022} found that SN\,2020tlf exhibited pre-explosion activity that persisted from 130 days prior the terminal explosion, subsequent flash-ionization features were observed. \citet{Jacobson-Galan_2022} found that the progenitor of SN\,2020tlf had a mass loss rate of $\sim$\,10$^{-2}$\,M$_\odot$\,yr$^{-1}$, which those authors suggest may be consistent with nuclear flashes \citep[e.g.][]{Woosley_2015} or gravity-wave driven outbursts \citep[potentially creating as much as 1\,M$_\odot$ of ejected material, contributing to the CSM;][]{Quataert_2012,Wu_2021}.

 Early time photometric and spectroscopic observations of SN\,2023ixf suggest that there was mass loss prior to the terminal SN explosion. The RSG models utilized by \citet{Jacobson-Galan_2023_ixf} suggest that the progenitor underwent a super-wind mass loss phase, with a mass loss rate of $\sim$\,10$^{-2}$\,M$_\odot$\,yr$^{-1}$ for 3\,--\,6 years prior to the explosion. This mass loss created a confined CSM with a density of 10$^{-12}$\,g\,cm$^{-3}$ at a radius of 10$^{14}$\,cm, with the radial extent of the CSM being 0.5\,--\,1.0\,$\times$\,10$^{15}$\,cm. \citet{Hosseinzadeh_2023_ixf} presented an analysis of the early-time light curve of SN\,2023ixf, finding that after the first day post-discovery, the light curve deviates from a power law or shock-cooling models, suggesting that this could be explained by precursor activity. \citet{Grefenstette_2023_xray_23ixf} report hard X-ray spectral observations of SN\,2023ixf from \textit{NuSTAR} consistent with a confined CSM with radial extent $\textless$\,10$^{15}$\,cm and progenitor mass loss rate of $\sim$\,3\,$\times$\,10$^{-4}$\,M$_\odot$\,yr$^{-1}$. \citet{2023arXiv230813101P} found that $Swift$ did not detect soft X-ray emission from SN\,2023ixf until $\sim$3 days post-explosion and concluded that the mass loss rate of the progenitor was $\lesssim$\,5\,$\times$\,10$^{-4}$\,M$_\odot$\,yr$^{-1}$ with a CSM radius of $\sim$\,4\,$\times$\,10$^{15}$\,cm and also that the CSM was asymmetric. Furthermore, using the Sub-Millimeter Array, \citet{Berger_2023} placed constraints on the CSM extent of $\sim$\,2\,$\times$\,10$^{15}$\,cm and pre-SN mass loss rate of $\sim$\,10$^{-2}$\,M$_\odot$\,yr$^{-1}$. Those authors also suggest that the  CSM was inhomogenous, possibly explaining the inconsistent mass loss rate from X-ray observations.

 Due to the proximity of SN\,2023ixf and the subsequent CSM interaction elucidated from early-time observations, it is a prime target for investigations into pre-SN activity. Indeed, several studies have already explored pre-explosion light curves for pre-SN outbursts. When considering pre-explosion \textit{Spitzer} data, \citet{Kilpatrick_2023_ixf} noted that the progenitor was detected at 3.6\,$\mu$m and 4.5\,$\mu$m. These infrared (IR) detections spanned between MJD\,53072\,--\,58781 and displayed variability with brightenings of $\sim$\,10\,$\mu$Jy with a periodicity of around 1000 days. \citet{Kilpatrick_2023_ixf} interpret this variability as being consistent with $\kappa$-mechanism oscillations \citep[opacity-driven variability;][]{Li_1994, Heger_1997, Paxton_2013}. \citet{Jencson_2023_ixf} also presented the \textit{Spitzer} photometry along with ground-based $J$ and $K_s$-band data spanning 13 years, up to 10 days before the SN explosion. These authors found that SED fits to the IR data suggest a luminous, dusty RSG progenitor with a luminosity of $\log\left(L/L_\odot\right)$\,=\,5.1\,$\pm$\,0.2 and a temperature of 3500$^{+800}_{-1400}$\,K and a mass loss rate of 3\,$\times$\,10$^{-4}$\,--\,3\,$\times$\,10$^{-3}$\,M$_\odot$\,yr$^{-1}$. Similarly, \citet{Soraisam_2023_ixf} found, using both the \textit{Spitzer} and ground-based $JHK$ data, a progenitor with  $\log\left(L/L_\odot\right)$\,=\,5.27\,$\pm$\,0.12 at T\,=\,3200\,K or $\log\left(L/L_\odot\right)$\,=\,5.37\,$\pm$\,0.12 at T\,=\,3500\,K corresponding to a progenitor mass of 20\,$\pm$4\,M$_\odot$. These findings indicate that the progenitor of SN\,2023ixf is fairly luminous compared to previously observed RSG SN progenitors, suggesting a massive RSG \citep[e.g.,][]{Smartt_2015}. Using archival Galaxy Evolution Explorer ({\it GALEX}) data, \citet{Flinner_2023} explore the near and far UV activity of the progenitor of SN\,2023ixf up to 20 years prior to the explosion, finding no outbursts in the UV to limits of L$_{\mathrm{NUV}}$\,=\,1000\,L$_\odot$ and L$_{\mathrm{FUV}}$\,=\,2000\,L$_\odot$. \citet{Dong_2023_ixf} investigate the pre-SN photometry obtained with the Zwicky Transient Facility (ZTF), the Asteroid Terrestrial-impact Last Alert System (ATLAS) and DLT40. While these data did not reveal any outbursts, \citet{Dong_2023_ixf} incorporated the pre-SN outburst models presented by \citet{Tsuna_2023} in order to put constraints on pre-SN activity. Those authors found that a precursor event with peak M$_r$\,=\,--9 would have had a duration of less than 100 days, while an outburst with M$_r$\,=\,--8 must have had a duration of 200 days or less. They suggest that an outburst similar to the models of \citet{Tsuna_2023} or what was seen prior to SN\,2020tlf was not likely to have occurred in SN\,2023ixf. Though SN\,2023ixf may not have suffered large outburst-like events, the confined CSM \citep[for example, see][who found that the CSM was close to the progenitor]{2023arXiv230813101P} must have originated from some enhanced mass-loss mechanism. Furthermore, \citet{Neustadt_2023_ixf} used archival data from the Large Binocular Telescope spanning 5,600\,--\,400 days prior to SN\,2023ixf to search for optical variability. Those authors found that there was no $R$-band variability to the 10$^3$\,L$_\odot$ level in the time frame of these data. \citet{2023arXiv230813101P} explored optical and X-ray pre-explosion data from ATLAS, ZTF, the All-Sky Automated Search for Supernovae (ASAS-SN), \textit{Swift}, \textit{XMM-Newton} and \textit{Chandra}, finding no pre-explosion variability and constrain any optical pre-SN outburst to $\lesssim$\,7\,$\times$\,10$^4$\,L$_\odot$ and X-ray pre-SN outburst to a limit of $\sim$\,6\,$\times$\,10$^{36}$\,erg\,s$^{-1}$. 

In this work, we present long-baseline pre-explosion photometric data of SN\,2023ixf spanning $\sim$\,5,000 days to a few days before the SN from Pan-STARRS in \textit{grizy} bands and also multi-year stacks in \textit{wizy} bands. These data were obtained through the Young Supernova Experiment \citep[YSE;][]{Jones_2021}. We analyze these data in search of pre-SN outbursts whose presence may be indicated by the already observed CSM interaction and variability in the IR. In section~\ref{sec:phot} we describe our methodology to systematically search for pre-explosion detections within the Pan-STARRS data. In section~\ref{sec:results} we will discuss the findings from our long baseline pre-explosion limits and make comparisons to known pre-SN outbursts. We combine these results with consolidated data from the existing literature to model the progenitor spectral energy distribution in section\,\ref{sec:prog}. In section\,\ref{sec:net}, we describe our method for using a pre-SN outburst model to train a multilayer perceptron classifier in order to search for pre-SN outbursts. We then use these models to constrain possible outburst properties. We repeat the SED analysis and neural net methodology to probe for possible variability of the progenitor prior to the SN explosion in section\,\ref{sec:vary}. Finally, we analyze the host in section\,\ref{sec:host} in terms of the spatial association of SN\,2023ixf with star formation. We conclude in section~\ref{sec:conc}.

\section{Photometry} \label{sec:phot}

We present pre-explosion data for SN\,2023ixf from Pan-STARRS \citep{Chambers_2016}. Pan-STARRS is comprised of a duo of 1.8\,m telescopes, PS1 and PS2, near the peak of Haleakala on the island of Maui. These data span from 19 Jan 2010\,--\,12 May 2023, using \textit{gwrizy} filter sets \citep{PS1}. In total, there are 313 PS1 pre-SN photometric observations over a 4,851 day baseline. These have a typical depth of 20.4 averaged over all \textit{grizy} filters. In the following, we present a custom pipeline to carefully measure the limiting magnitude of each individual exposure.

\begin{deluxetable}{ll|l|l|l|l}
\caption{Pre-explosion Pan-STARRS 80\% detection confidence limits in \textit{gwrizy} filters. A complete version of this table in machine-readable format is available online.\label{tab:data}}
\tablecolumns{6}
\tablehead{
\colhead{Type} &
\colhead{Phase (days)} & \colhead{MJD} & \colhead{Filter} & \colhead{Lim. Mag.} & \colhead{\# Aps.}  }
 \startdata\hline
Single & -4040.393 & 56042.44 & g & 22.20 & 12 \\ 
Single & -4040.383 & 56042.45 & g & 22.24 & 11 \\ 
Single & -3687.433 & 56395.40 & g & 22.00 & 12 \\ 
Single & -3014.173 & 57068.66 & g & 21.84 & 10 \\ \hline
Stack & -- & -- & $w$ & 24.80& -- \\ 
Stack & -- & -- & $i$ & 23.80& -- \\ 
Stack & -- & -- & $z$ & 23.00& -- \\ 
Stack & -- & -- & $y$ & 20.03& -- \\ 
\enddata
\end{deluxetable}

\subsection{Pre-supernova eruption detection pipeline} \label{sec:pipe}

We measure the pre-explosion photometry using {\tt Photpipe} \citep{Rest+05} to ensure highly accurate photometric measurements and to account for pixel-to-pixel correlations in the difference images and host galaxy noise at the SN location. {\tt Photpipe} is a well-tested pipeline for measuring SN photometry and has been used to perform accurate measurements from Pan-STARRS in a number of previous studies (e.g., \citealt{Rest14,Foley18,Jones18,Scolnic18,Jones19}). In brief, {\tt Photpipe} takes as input Pan-STARRS images which have been reduced by an initial image processing pipeline. Our pre-processing pipeline resamples the images and astrometrically aligns them to match skycells in the Pan-STARRS 1 (PS1) sky tessellation. Geometric distortion is then removed. We then measure image zero points using {\tt DoPhot} \citep{Schechter+93} to measure the photometry of stars in the image and comparing to stars in the PS1 Data Release 2 catalog \citep{Flewelling+16}. {\tt Photpipe} then convolves a template image from the PS1 3$\pi$ survey \citep{Chambers2017},with data taken between the years 2010 and 2014, using a kernel that consists of three superimposed Gaussian functions. This kernel is designed (and fit)  to match the point spread function (PSF) of the survey image. We then subtract the template from the science image using {\tt hotpants} \citep{Becker15}.  Finally, {\tt Photpipe} uses {\tt DoPhot} to measure fixed-position (i.e., forced) photometry of the SN at the weighted average of its location across all images.  Further details regarding this procedure are given in \cite{Rest14} and \cite{Jones19}.

To account for underlying structure in the bright host galaxy of SN~2023ixf, which could cause larger-than-expected pre-explosion photometric noise in the difference image \citep{Kessler15,Doctor17,Jones17}, we forward model our full reduction pipeline. We simulate a noisy detection by estimating the signal-to-noise that would be recovered from a source of a given flux assuming the following sources of uncertainty: (1) the Poisson noise at the SN location (i.e., the square root of the counts) and (2) Gaussian noise from the background (i.e., the standard deviation of flux values measured from random difference-image apertures at coordinates with approximately the same underlying host galaxy surface brightness as exists at the SN location. The apertures used in our reduction pipeline must closely match the background noise statistics at the site of SN\,2023ixf in order to obtain a more rigorous calculation of our detection limits. In order to select these apertures, a grid of $3^{\prime\prime}$ apertures are placed over the host in images in each \textit{grizy} filter. The aperture grid, (with 367 trial apertures) is placed over a $57^{\prime\prime}$\,$\times$\,$57^{\prime\prime}$ area (covering the host region in the images), with no overlap between apertures. An aperture is also placed over the location of SN\,2023ixf \citep[determined using the coordinates of SN\,2023ixf from][]{Kilpatrick_2023_ixf}. The distribution of the flux values within the aperture containing SN\,2023ixf is measured and then compared with the flux distributions of the apertures in the grid. Apertures from the grid are then chosen for use in our source injection method. These apertures are selected using a given flux distribution similarity tolerance (here our tolerance was chosen such that at least ten apertures are found in each image) on the distribution of parameters. More specifically, we select apertures based on the mean (within 25\% of the standard deviation of the mean), standard deviation (within 10\% of the standard deviation), the skew (within 10\%  of the skew) and kurtosis (within 10\% of the kurtosis) of the distribution. The number of apertures differ per filter and these apertures largely follow the spiral arms of the host, similar to the location of SN\,2023ixf. A summary of these data (including the number of apertures found in each image) found from PS1 is tabulated in Table\,\ref{tab:data}.

To search for pre-SN emission in all Pan-STARRS images, we perform an idealized fake source injection within each chosen aperture to estimate the recovery fraction (i.e., the fraction of apertures where the injected source is recovered at $>3\sigma$ significance) as a function of the injected source flux. To find true pre-SN detections, we compared the derived limiting magnitude to photometric measurement from {\tt Photpipe} at the SN location. We label detections as real if the latter is brighter than the former. For each image, we estimate the limiting magnitude based on the flux (i.e., in analog-to-digital units, ADU, given the zero point calculated above) associated with an 80\% recovery fraction in the chosen background apertures. We consider this 80\% recovery fraction as a detection. At this recovery level, we do not generate false positive detections that would be statistically expected in more standard photometric methodologies. 

To test the validity of any possible detections, we perform a more robust fake source injection routine in the science images, also using {\tt Photpipe}, to estimate a new set of recovery curves for each epoch where there may be a possible detection.  This procedure is slower and more computationally intensive than the procedure described above and uses the PSF shape determined by {\tt DoPhot} \citep[i.e., a seven parameter Gaussian as described in][]{Schechter+93} to create artificial sources with a known flux and at the same aperture locations described above in the original science image.  We then repeat the reduction process, including image subtraction with {\tt hotpants}, in order to simulate the effect of convolution noise in the recovery of each source.  Finally, we perform forced photometry at the source location to simulate the detection of sources whose sky locations are known a priori and create recovery curves as a function of the injected source flux.  In order to obtain a statistically significant number of sources over a broad range of magnitudes, we repeat this process with the same image and aperture locations until we have forced photometry for 1,500 sources from 17--24~mag.  Here we also adopt the 80\% recovery fraction as the limiting magnitude, which we then compare to the photometric measurement at the SN location. 

\section{No evidence of pre-explosion activity in Pan-STARRS data} \label{sec:results}

\begin{figure*}[!t]
	\includegraphics[trim={0 2cm 0 0},clip,width=0.99\textwidth]{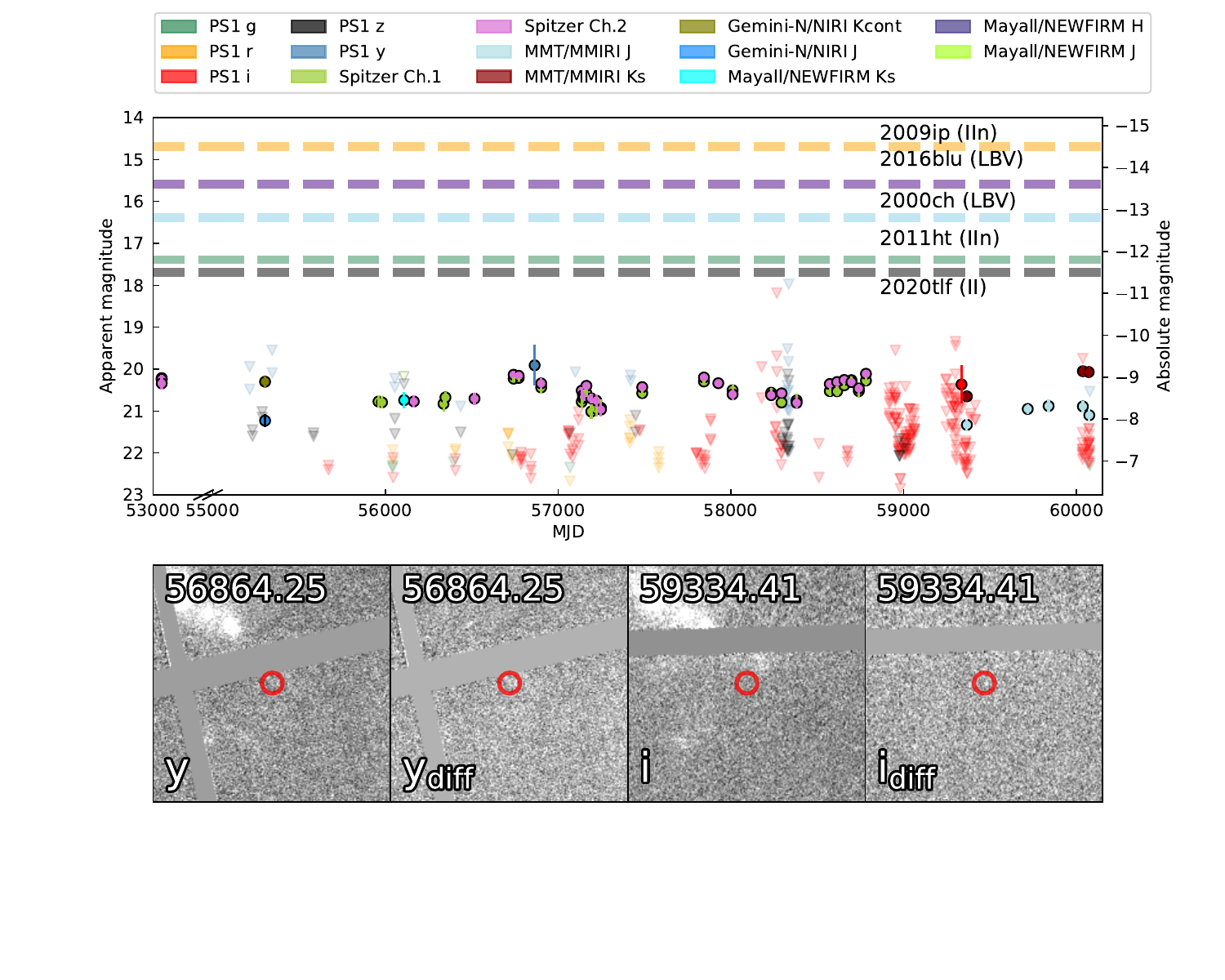}
    \caption{(\textit{Upper:}) The long-baseline pre-explosion light curve of SN\,2023ixf. We present PS1 \textit{grizy} photometry spanning 4851 to 6 days prior to SN\,2023ixf. \textit{Spitzer}, \textit{Gemini/NIRI}, \textit{MMT/MMIRI} and \textit{Mayall/NEWFIRM} data (all detections, originally presented in \citealt{Jencson_2023_ixf, Kilpatrick_2023_ixf}) are also shown. The Pan-STARRS limits are from our source injection method. For the Pan-STARRS data, the two instances of the source injection finding possible sources at the 50\% recovery fraction limit (in $i$- and $y$-bands) are marked as a circle and 80\% recovery fraction limits are shown as a downward pointing triangle. These two sources, however, are detected at a signal-to-noise level $\textless$\,3$\sigma$. Overplotted are also the peak absolute magnitude of pre-SN outbursts of three other transients, SN\,2020tlf  \citep[a luminous SN\,II, bolometric peak luminosity][]{Jacobson-Galan_2022}, SN\,2011ht \citep[a SN\,IIn with a plateau light curve;][whose $z$-band peak absolute magnitude shown]{Fraser_2013_11ht}, SN\,2009ip \citep[a well studied SN\,IIn with bright pre-cursor eruptions][peak visual magnitude shown]{Mauerhan13a}, SN\,2000ch and AT\,2016blu \citep[SN impostors][]{Aghankhanloo_2023_16blu, Aghankhanloo_2023_2000ch, Pastorello_2010}. (\textit{Lower:}) PS1 cutouts of the epochs of the possible sub-3$\sigma$ detections in the $i$ and $y$-band, both the science images (panels 1 and 3) and difference images (panels 2 and 4 are shown. The location of the transient is marked by a red circle. There is no visible detection in these images.}
    \label{fig:grizy}
\end{figure*}

We present the Pan-STARRS long baseline \textit{grizy} light curve in Figure~\ref{fig:grizy}. Through our 4,851 day pre-explosion baseline, we find no detections at the 80\% aperture recovery fraction in the $g$, $r$, $i$, $z$ or $y$ bands. The median limits we found in each filter are 22.0\,mag in the $g$ band, 21.6\,mag in the $r$ band, 21.3\,mag in the $i$ band, 21.3\,mag in the $z$ and 20.1 in the $y$ band, or: $M_g\,=\,-7.2$\,mag; $\,M_r=\,-7.6$\,mag; $M_i\,=\,-7.9$\,mag; $M_z\,=\,-7.9$\,mag and $M_y\,=\,-9.1$\,mag. While these source injection limits are obtained using difference images, the templates used to make the difference images are $\sim$2\,--\,3 mags deeper than the individual epoch images at the same position so our measurements are sensitive to the depth of the single epoch science images. This implies that the measurements from our difference images between the individual images and the template images are limited by the depth of the individual images. Therefore, the underlying progenitor flux in the template image is insignificant when measuring limits on outburst luminosity in difference images. The range of literature progenitor bolometric RSG luminosities is $\sim$\,10$^{4.39\,-\,5.52}$\,L$_\odot$, corresponding to absolute magnitudes of  $\sim$\,--6.2 to --9.0 \citep{Davies_2020_RSG}, with the most luminous known RSG being UY\,Scuti, with $\log\left(L/L_\odot\right)$\,$\approx$\,5.52 \citep{Arroyo-Torres_2013}. Our limits are therefore mostly on the upper end of, or are brighter than the range of the bolometric luminosities of observed RSG SN progenitors. 

We obtain multi-year stacks in the $w$, $i$, $z$ and $y$ filters to probe for progenitor detections. These data were compiled using the data from the Pan-STARRS Survey for Transients (PSST), which itself uses the $wiz$ data from near-Earth object searches \citep{Huber_2015}. As the $w$ filter does not contain color information, it is not used by YSE. Rather, these data are from coincidental observations with YSE fields (and are therefore not included in light curve analysis). Forced photometry of these non-difference imaged stacks reveal that there are \textbf{no} progenitor detections to limits (3$\sigma$ limits) of 24.80 mag in the $w$ band, 23.80 mag in the $i$ band, 23.00 mag in the $z$ band and 20.03 mag in the $y$ band, or $M_w$\,=\,--4.4\,mag, $M_i$\,=\,--5.4\,mag, $M_z$\,=\,--6.2\,mag, $M_y$\,=\,--9.2\,mag.

There is weak evidence of possible detections in the $i$- and $y$-bands at MJD\,59334.41 and 56864.25 (--753.6 and --3223.8 days relatively to explosion), respectively, at a less stringent 50\% recovery limit; however, these are \textit{not} detections at the 80\% limit. As these epochs only meet a 50\% recovery fraction, we inspect these epochs in more detail. The $i$-band detection is at a $\sim$\,2.4$\sigma$ detection significance, while the $y$-band detection is at $\sim$\,2.2$\sigma$ detection significance with these being single images. We present cut-out images of these detections in Figure~\,\ref{fig:grizy}. There are no clear visible sources at the location of SN~2023ixf in the thumbnails, consistent with our low significance detections. Therefore we consider these as non-detections. For our 313 Pan-STARRS observations, one would expect $\sim$\,15 observations at the 2\,$\sigma$ level and $\sim$\,1 observation at the 3\,$\sigma$ level false-positive detections if using a more standard photometric method. Our source injection method produces no 2\,$\sigma$ or 3\,$\sigma$ detections at the 80\% recovery fraction.

Finally, we compare our long-baseline pre-explosion light curve to previously identified precursor outburst events in other SNe. Firstly, SN\,II, 2020tlf had precursor outbursts that peaked at an absolute magnitude $\sim$\,--11.5 \citep{Jacobson-Galan_2022}. As shown in Figure\,\ref{fig:grizy}, all of our PS1 limits are deeper than SN\,2020tlf-like pre-SN outbursts, obtained with a similar method to this work. To compare to the SNe\,IIn pre-SN outbursts found in the literature, we select two SNe\,IIn which are examples of the upper and lower luminosity ranges of observed SN\,IIn precursor outbursts \citep[e.g.][]{Strotjohann_2020}\footnote{Pre-explosion outbursts in SNe\,IIn are perhaps the best known, e.g. between 2018 and 2020, 18 SNe\,IIn observed with ZTF were found to have precursor events.}. At the fainter end there is SN\,2011ht, where \citet{Fraser_2013_11ht} report an outburst a year before the SN event, with it peaking at an absolute magnitude $\sim$\,--11.8. On the brighter end of the SN\,IIn precursor eruption scale, there is SN\,2009ip. Initially discovered as an ``impostor", SN\,2009ip likely suffered its terminal explosion in 2012, with the 2009 eruption peaking at an absolute magnitude of $\sim$\,--14.5. However, the nature of SN\,2009ip is still a topic of debate \citep[see][]{Berger_2009,Miller_2009,Smith_2010,Drake_2010, Foley_2011,Pastorello_2013,Margutti_2014,Smith_2014_2009ip, Mauerhan13a, Pessi_2023}. Our limits and the progenitor detections of SN\,2023ixf are dimmer than the outbursts seen in the RSG progenitor of SN\,2020tlf by at least 2.5\,mag and are much fainter than the possibly LBV-like outbursts seen prior to some SNe\,IIn such as SN\,2009ip. In addition to the pre-SN explosions associated with these SNe\,IIn, we can also compare to some SN impostors, many of which are also interpreted as eruptions of LBV-like progenitors. For example, SN\,2000ch and AT\,2016blu are both SN impostors with ongoing observed activity \citep{Aghankhanloo_2023_16blu, Aghankhanloo_2023_2000ch, Pastorello_2010}. SN\,2000ch peaked at an absolute magnitude of $\sim\,$--12.8 and AT\,2016blu peaked at $\sim$\,--13.6 (lying in between the pre-SN outburst in the SNe\,IIn range).

\section{Progenitor Analysis via Stacked Data} \label{sec:prog}

To constrain the properties of the progenitor of SN\,2023ixf, spectral energy distributions (SEDs) of the progenitor are presented by a number of authors \citep[e.g.][]{Kilpatrick_2023_ixf, Jencson_2023_ixf, Niu_2023, Neustadt_2023_ixf, Xiang_2023, Soraisam_2023_ixf}. Detections in \textit{Spitzer} channel 1 and channel 2, MMT $J$, Gemini/NIRI $J$, UKIRT $JHK$ and \textit{Hubble Space Telescope} (\textit{HST}) $F814W$ and $F675W$ are used here. As stated, pre-SN observations (particularly those from \textit{Spitzer}) reveal a highly variable progenitor in the decade up to SN. We must account for the scatter in reported photometric measurements, and also the variability in the IR data. As our mean estimate in each band, we take an average of these flux measurements over independent measurements and time. For the uncertainties on these measurements, we account for two contributions: the systematic scatter in reported measurements of the \textit{same} observations, and intrinsic variability. In the latter case, we use the range of reported AB magnitudes as an estimate for the systematic uncertainty where the error interval is the range of values per filter, and in the case where epochs have multiple measurements, we add the average scatter per epochs in quadrature\footnote{This table is provided as a github repository at \url{https://github.com/AstroSkip/pre_sn_23ixf.git}.}. 

We use the radiative transfer code \texttt{DUSTY} \citep{Kochanek_2012_dusty} to constrain the progenitor properties. Following \cite{Kochanek_2012} and \cite{Kilpatrick_2023_ixf}, we use the Model Atmospheres with a Radiative and Convective Scheme (MARCS) grid of RSG spectra \citep[e.g.][]{Gustafsson_1975, Gustafsson_2008} as an internal heating source within an spherically symmetric shell of dust. We note that, while the immediate CSM showed signs of asymmetry, \texttt{DUSTY} assumes a spherically symmetric dust shell. MARCS provides a grid of $15$\,M$_\odot$ RSG spectra, with varying temperatures, surface gravities and metallicities. Here, we explore solar metallicity models with $\log(g)=0$ and progenitor effective temperatures between the range 3300\,K and 4500\,K. The MARCS models are then used as internal heating sources for the \texttt{DUSTY} models, allowing us to estimate the dust properties of the progenitor system. We specifically vary the optical depth of the dust ($\tau_V\in{0,10}$), the ratio of the outer to inner radii of the dust shell ($\log_{10}(R_\mathrm{out}/R_\mathrm{in})\in{2,4})$, and the inner temperature of the dust ($T\in{10,1000\,K}$). We test carbonaceous and silicate dust models, as dust of both types is commonly seen. Finally, we fit for luminosity between $\log\left(L/L_\odot\right)$\,=\,3\,--\,6. 

We generate an interpolated grid of pre-computed \texttt{DUSTY+}MARCS models and use the Bayesian nested sampling algorithm \texttt{Dynesty} \citep{Speagle_2020} to constrain the progenitor properties. We additionally fit for an extra white-noise term, $\sigma^2$, to capture systematic uncertainties which may be underrepresented in our measurements, i.e. a parameter that represents the fractional underestimate of the uncertainties in log-space. From the posterior distributions, we infer the following values for the progenitor luminosity with carbon based dust (graphitic): a luminosity of $\log\left(L/L_\odot\right)$\,=\,5.12$^{+0.15}_{-0.21}$, an optical depth $\tau$\,=\,8.23$^{+0.90}_{-1.20}$, an RSG temperature of 3935$^{+335}_{-296}$\,K, a dust temperature of 405$^{+276}_{-268}$\,K, a $\log_{10}(R_\mathrm{out}/R_\mathrm{in})$ of 3.10$^{+0.59}_{-0.71}$, and $\sigma$ of --5.75$^{+2.92}_{-2.89}$. Our low value of $\sigma$ suggests that we do not significantly underestimate uncertainties. Silicate dust models were trialed and were not as good a fit to the data as the graphitic dust, with reduced $\chi^{2}$ values of 1.8 for silicate dust and 0.6 for graphitic dust. Therefore, we only consider the graphitic dust models. These values are broadly consistent with previous studies on the progenitor of SN\,2023ixf. Our luminosity is consistent with most other work within the uncertainties \citep{Jencson_2023_ixf, Niu_2023, Qin_2023,VanDyk_2023, Soraisam_2023_ixf, Neustadt_2023_ixf, Xiang_2023}, with \citet{Soraisam_2023_ixf} finding the highest luminosity at $\log\left(L/L_\odot\right)$\,=\,5.27$\pm$0.12 or $\log\left(L/L_\odot\right)$\,=\,5.37$\pm$0.12 dependent on the temperature used in their fits. Our RSG temperature is on the higher end of the range from other studies, with \citet{Kilpatrick_2023_ixf} finding the next hottest temperature at 3920$^{+200}_{-160}$\,K, but also our uncertainties are larger due to the scatter in the photometry. However, our temperature is consistent with a number of the studies within uncertainties \citep{ Niu_2023,VanDyk_2023, Neustadt_2023_ixf, Jencson_2023_ixf}.

\begin{figure}[!t]
	\includegraphics[width=0.99\columnwidth]{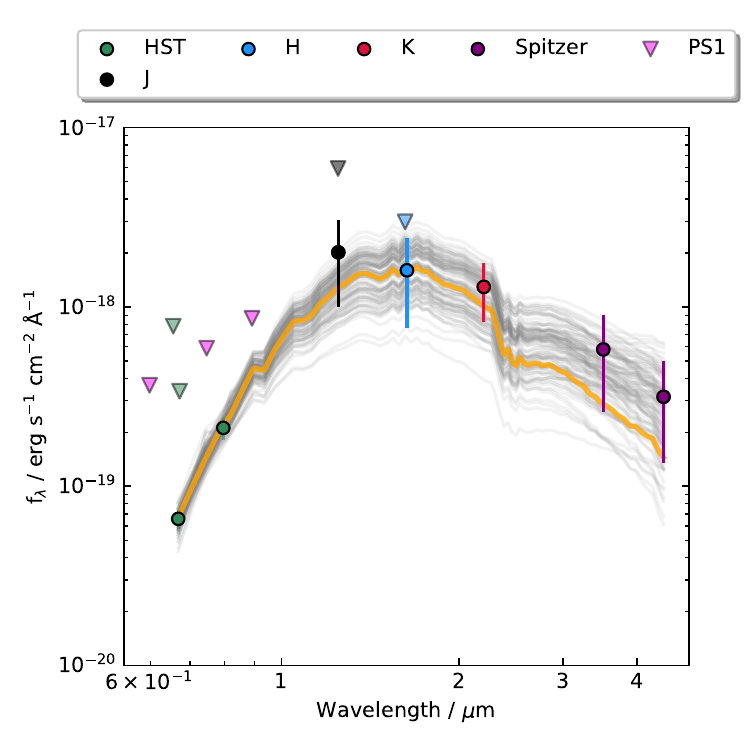}
    \caption{Consolidated photometry of the progenitor from the SN\,2023ixf literature and our best-fit models. We use the photometric measurements presented by \citet{Kilpatrick_2023_ixf}, \citet{Jencson_2023_ixf}, \citet{Soraisam_2023_ixf}, \citet{Xiang_2023} and \citet{Niu_2023}. These data consist of \textit{Spitzer} channel 1 and channel 2, MMT $J$ and $Ks$, Gemini/NIRI $J$, UKIRT $HJK$, Mayall/NEWFIRM $Ks$ and \textit{HST} $F814W$ and $F675W$. The model SED that represents the median posterior values is plotted in orange and random draws are plotted in gray for reference.}
    \label{fig:sed}
\end{figure}

\begin{figure}[!t]
	\includegraphics[width=0.99\columnwidth]{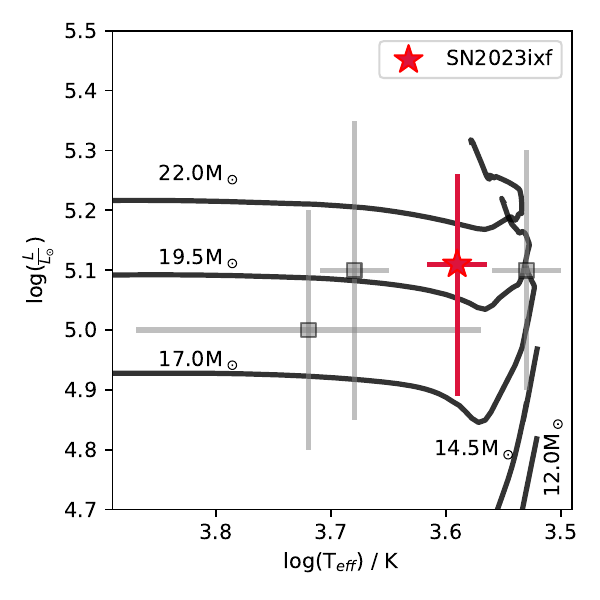}
    \caption{Evolutionary tracks from MIST compared with our progenitor measurements from our SED fits. The red star is the fit for SN\,2023ixf. The gray squares are RSG SN progenitors from \citet{Smartt_2015}.}
    \label{fig:evo}
\end{figure}

Our SED fits are presented in Figure\,\ref{fig:sed}. In addition to detections of the presumed progenitor, we also plot limits from the Pan-STARRS $wizy$ multi-year stacks and limits from $H$-band (MJD 56108) and $J$-band (MJD 56107). These limits are consistent with our SED fits. We note that progenitor detections that are at single epochs are $HST$ F814W (MJD\,52594) and F675W (MJD\,51261). Our SED fits are consistent with most (but not all) of the previous literature \citep[see summary by][]{Qin_2023}. Finally, we compare our SED fits to the MESA Isochrones \& Stellar Tracks (MIST) evolutionary models \citep{Dotter_2016, Choi_2016} assuming a non-rotating star and solar metallicity models. We consider models to be consistent if their final \textit{luminosity} is consistent with our derived values. Assuming a graphitic dust model, we find that our progenitor properties are consistent with a  14\,--\,20\,M$_\odot$ star (see Figure\,\ref{fig:evo}). This mass range is too high for the electron-capture scenario suggested by \citet{Xiang_2023}. 

 In our SED, the largest scatter is in the $H$-band from data presented by \citet{Soraisam_2023_ixf} with an uncertainty of $\sim$\,1 mag, this is due to the variability of the progenitor in the IR. Furthermore, the reported \textit{Spitzer} data has a large scatter in both the 3.6\,$\mu m$ and 4.5\,$\mu m$ channels with the range in the average brightness being 0.91\,mag and 0.72\,mag respectively. Other methodological differences such as differences in SED models have an effect on calculated progenitor parameters. For example, \citet{Soraisam_2023_ixf} use a RSG period-luminosity relation to obtain their high luminosities. Others phase average their data to account for variability \citep{Jencson_2023_ixf}, while others assume no variability when creating inputs for their SEDs \citep{Kilpatrick_2023_ixf}. \citet{VanDyk_2023} incorporated the variability in the IR using the range in the IR measurements and models of the $J$-band to $V$-band variability to estimate an uncertainty \citep{Smith_2002_Mira, Riebel_2012}. \citet{Niu_2023} add a 0.5 magnitude uncertainty to their optical measurements to account for variability. Furthermore the dust models used differ, with some using carbon based (graphitic) dust models \citep{Kilpatrick_2023_ixf, Niu_2023} and others using silicate based dust models \citep{Jencson_2023_ixf, VanDyk_2023}. 
 
\begin{figure}[!t]
	\includegraphics[width=\columnwidth]{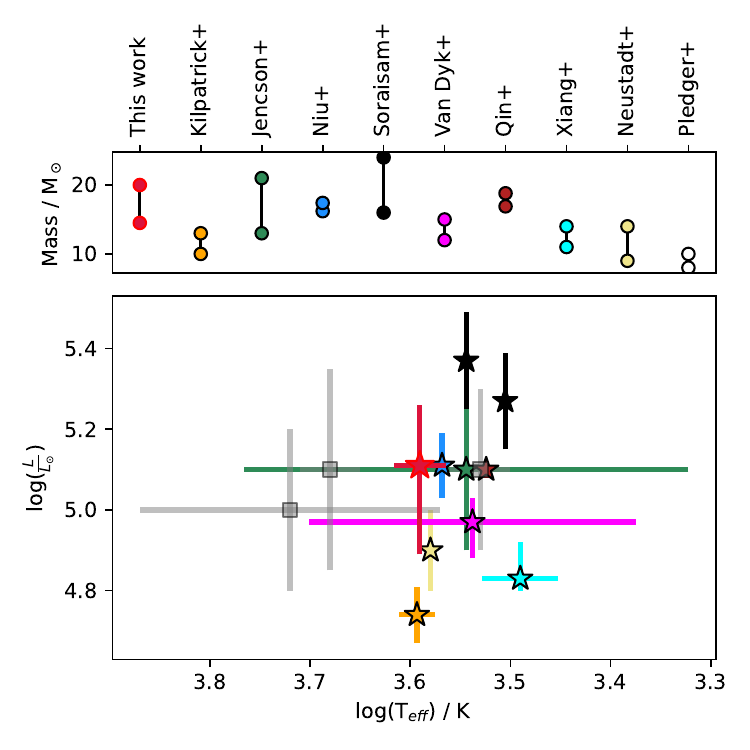}
    \caption{\textit{Top:} Comparison of our progenitor mass compared with values from previous work. \textit{Bottom:} Comparison of our progenitor luminosity and effective temperature compared with values from previous work. The gray squares are RSG SN progenitors from \citet{Smartt_2015}. }
    \label{fig:progscatter}
\end{figure}

Our progenitor mass estimates, as expected, lie within the range of reported values (which shows substantial scatter). The range of reported progenitor masses includes the lowest end of the range for CCSN progenitors--\citet{Pledger_2023} reports a progenitor mass of 8\,--\,10\,M$\odot$, using isochrone fitting using \textit{HST} pre-explosion data. The SED analysis of \citet{Jencson_2023_ixf}, using the Grid of Red supergiant and Asymptotic Giant Branch ModelS \citep[GRAMS, with silicate dust;][]{Sargent_2011, Srinivasan_2011}, suggests an RSG with mass 17\,$\pm$\,4\,M$_\odot$, luminosity $\log\left(L/L_\odot\right)$\,=\,5.1\,$\pm$\,0.2 and RSG temperature of 3500$^{+800}_{-1400}$\,K. \citet{Niu_2023} also find a massive RSG progenitor with mass 16.2\,--\,17.4\,M$_\odot$ and luminosity $\log\left(L/L_\odot\right)$ = 5.11 for a model SED with graphitic dust and RSG temperature of 3700\,K. \citet{VanDyk_2023} used SED fitting which accounted for the variability of the progenitor and single-star stellar evolution models (GRAMS, with silicate dust) to constrain a progenitor with mass 12\,--\,15\,M$_\odot$. Here, they derived a luminosity of 7.6\,--\,10.8\,$\times$\,10$^4$\,L$_\odot$ with an RSG temperature of 3450$^{+250}_{-1080}$\,K, which they suggest is similar to the Galactic RSG, IRC\,--\,10414. \citet{Xiang_2023} use the \textit{HST} and \textit{Spitzer} data to fit an SED to a dusty RSG model, finding a very cool RSG temperature of 3090\,K, a progenitor mass of 12$^{+2}_{-1}$\,M$_\odot$ with $\log\left(L/L_\odot\right)$ = 4.8. \citet{Xiang_2023} also suggest that the IR colors of the progenitor of SN\,2023ixf may suggest a super-asymptotic giant branch star, in which case it would be on the lower end of the CCSN progenitor mass range of 8\,--\,10\,M$_\odot$ and possibly explode as an electron-capture SN. \citet{Qin_2023} use archival \textit{HST} data along with the \textit{Spitzer} data to infer a progenitor with mass 18$^{+0.7}_{-1.2}$\,M$_\odot$, a luminosity of $\log\left(L/L_\odot\right)$ = 5.1\,$\pm$\,0.02, and RSG temperature of 3343\,$\pm$\,26\,K. \citet{Neustadt_2023_ixf} infer a progenitor mass of 9\,--\,14\,M$_\odot$ from their data from the Large Binocular Telescope (LBT) and a silicate dust model, with luminosity $\log\left(L/L_\odot\right)$ = 4.8\,--\,5.0. Generally, the differences in reported values in the literature may be attributed to a variety of factors described above, such as differences in the photometric treatment of the archival imaging of the progenitor, different dust models, stellar evolution tracks and SED fitting methods (e.g. fixing the effective temperature). We have incorporated the available photometric measurements from the literature to construct our SED which is well sampled in wavelength space, albeit with our conservative uncertainty treatment accounting for both the IR variability and differences in reported values from literature. We summarize and compare these values with the literature in Figure\,\ref{fig:progscatter}.

\section{Searching for pre-supernova outbursts with a neural net classifier} \label{sec:net} 

We search for pre-explosion outbursts in the PS1 data using a multilayer perceptron classifier. Multilayer perceptrons are neural networks comprised of at least three layers (input, hidden and output) with neurons that are fully connected and use a non-linear activation function, such as a sigmoid. Multilayer perceptrons are commonly used as relatively lightweight and fast-to-train classifiers due to their utility in distinguishing between complex non-linear datasets.  We train our classifier on model light curves which have injected outbursts following a pre-SN outburst model. These light curves assume the same form as our pre-explosion Pan-STARRS data in terms of filters and epochs. For each real observation, in some filter, there will be a model observation in the same filter, with each of the model light curves having 313 observations consistent with our data.

\begin{figure*}[t]
	\includegraphics[width=0.99\textwidth]{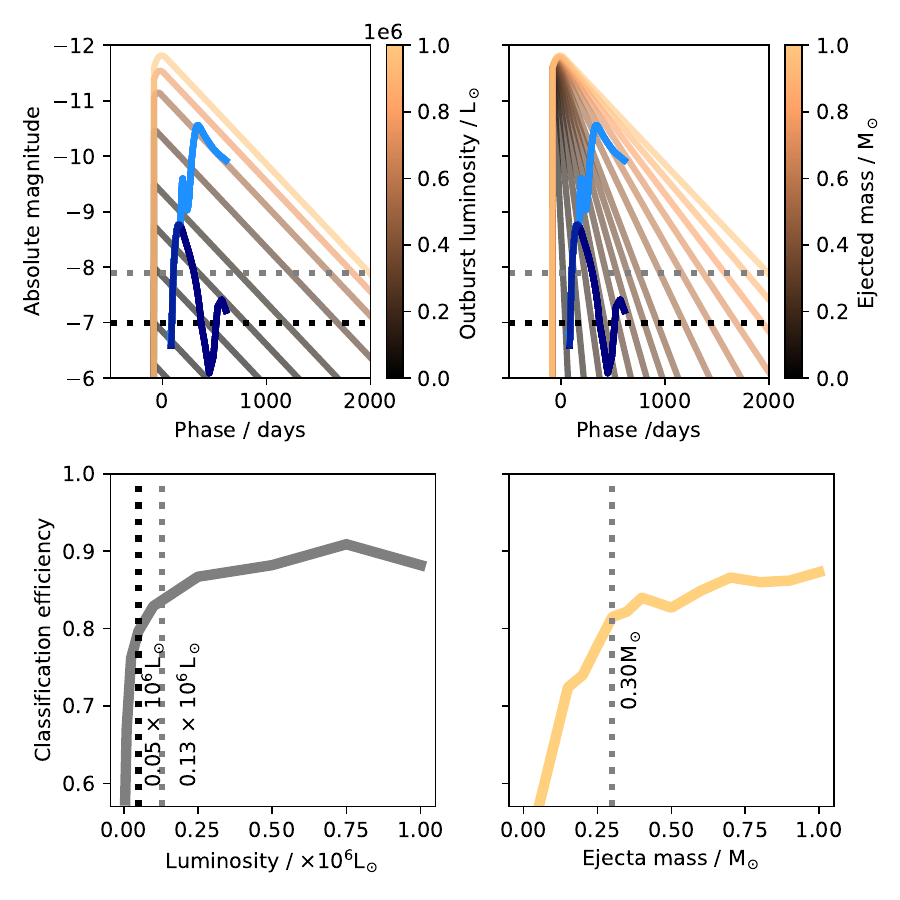}
    \caption{Detection efficiency curves from our neural network classifier. Using the test light curves described in section\,\ref{sec:net}, we can place constraints (defined as the limit at 80\% detection efficiency). On the top row, dotted lines are also plotted representing the luminosity threshold for the 80\% detection limit (black) and our averaged limits over \textit{grizy} filters (gray). On the top row, for comparison, in light blue is the double-long model from \citet{Tsuna_2023} and in dark blue is their single-small model. (\textit{Top left:}) example light curves for increasing injected luminosity. (\textit{Top right:}) example light curves for increasing ejecta mass. For the example light curve plots, only luminosity or ejecta mass were varied with other parameters fixed. The color-map transitions from brown to orange for higher injected luminosity/ejecta mass. 
    (\textit{Bottom left:}) detection efficiency curve for increasing injected luminosity. Dotted lines are also plotted representing the luminosity threshold for the 80\% detection limit (black) and our averaged limits over \textit{grizy} filters (gray). (\textit{Bottom right:}) detection efficiency curve for increasing ejecta mass. Overplotted is our upper ejected mass limit for SN\,2023ixf which is similar to the  mass loss estimate for SN\,2020tlf (gray dotted line) from \citet{Jacobson-Galan_2020}.}
    \label{fig:detect}
\end{figure*}

Our model takes the form of a blackbody SED expanding from the initial progenitor radius at a constant velocity, $v_\mathrm{ej}$, whose luminosity assumes no driving central power source \citep[e.g. recombination;][]{Arnett_1980, Villar_2017}:

\begin{equation}
    L = L_0 e^\frac{-(t-t_0)}{\tau_\mathrm{diff}}
\end{equation}
where $L_0$ is the initial input luminosity, $t_0$ is the time of eruption and $\tau_\mathrm{diff}$ is the diffusion time that takes the form:

\begin{equation}
    \tau_\mathrm{diff} = \frac{\kappa M_\mathrm{ej}}{\beta c R_0}
\end{equation}
where $c$ is the speed of light, $\kappa = 0.34 $ cm$^2$\,g$^{-1}$ is the opacity of H-rich material, $\beta=13.7$ is a geometric constant. $M_\mathrm{ej}$ (the ejecta mass) and $R_0$ (the progenitor radius) are free parameters of the model. We assume that the black body temperature self-consistently decreases until reaching $5000$\,K, at which point our photosphere begins to recede to maintain this temperature. Model realizations are shown in Figure~\ref{fig:detect}.

In this model, $L_0$, $R_0$, $t_0$, $v_{ej}$ and $M_{ej}$ are free parameters. In our training sets, we fix $v_{ej}$ to represent the measured wind velocity, high resolution spectroscopy indicates a wind velocity of $\sim$\,50\,km\,s$^{-1}$ \citep{Zhang_2023}. It should be noted that \citet{Smith_2023_ixf} found higher velocities that may originate from winds that have been radiatively accelerated. Our four free parameters are therefore the input luminosity, the pre-SN outburst time, progenitor radius, and the ejecta mass. 
We uniformly sample from a range of parameter values. We generate 10$^4$ training set light curves which are set at the distance of the host, M\,101 (6.9\,Mpc) and dust extinction is added \citep[with $A_V = 4.6$\, mag as per][]{Kilpatrick_2023_ixf} and with $R_V\,=\,3.1$ following the extinction law of \citet{Schlafly_2011}. 

Training sets are generated such that the resultant simulated light curves have observations at identical epochs at identical filters as the real data in the long baseline pre-explosion Pan-STARRS \textit{grizy} light curve. The uncertainties on these model observations are calculated by interpolating the uncertainties from flux-uncertainty maps from our source injection method described in section\,\ref{sec:phot}. In our model, we vary the input luminosity between 0\,--\,10$^6$\,L$_\odot$ with the maximum being chosen as it is of the order of the outburst observed in SN\,2020tlf \citep{Jacobson-Galan_2022}. We vary the ejecta mass uniformly and randomly between 0.01\,--\,1.00\,M$_\odot$, typical of pre-SN outbursts in the time frame that the CSM around SN\,2023ixf was formed \citep[e.g.,][]{Smith_2014}. The time of the injected eruption spans the time phase-space of our pre-explosion data. 

When sampling these model light curves to generate our training light curves, we convolve these pre-SN outbursts with the filter response curves for each of our \textit{grizy} filters in order to create a model observation. The filter response curves were obtained from the Spanish Virtual Observatory Filter Profile Service\footnote{http://svo2.cab.inta-csic.es/theory/fps/}. Furthermore, we illustrate how increasing the injected luminosity or ejecta mass has on the outburst light curves on the bottom two panels of Figure\,\ref{fig:detect}. These example light curves show the same increments in luminosity and ejecta mass with arbitrarily chosen ``middle of the range'' parameters fixed. This includes: a progenitor radius of 500\,R$_\odot$, an injected luminosity of 1.0\,$\times$\,10$^6$\,L$_\odot$ and an ejecta mass of 0.5\,M$_\odot$. 

 We use a multilayer perceptron in order to detect pre-SN eruptions within our PS1 light curve with 3 layers and 12 neurons in the first layer, 8 in the second and 1 in the third, using a combination of the standard sigmoid and relu activation functions. We train 2,500 epochs using the standard \texttt{adam} optimizer \citep{Kingma_2014}. After training our neural network to classify the presence of a pre-SN eruption (with an accuracy of $\sim$\,94\%), we then used the trained neural network to determine if such an eruption is present in the long-baseline pre-explosion \textit{grizy} Pan-STARRS light curve. Our neural net classifies these pre-explosion data as being consistent with there being \textit{no} detectable pre-SN outbursts in this 4,851 day range. 
 
 Given this non-detection, we place limits on the possible eruption models ruled out from our observations. We generate a test set of $\sim10,000$ eruptive light curves of various luminosities and ejecta masses and test the detection efficiency of our classifier. These parameters are increased incrementally (between 0\,--\,10$^6$\,L$_\odot$, and 0.01\,--\,1.00\,M$_\odot$). This is shown in Figure\,\ref{fig:detect}.

 With our parameter range, we can put a constraint on the injected luminosity of a pre-explosion outburst to be $\textless$\,5\,$\times$\,10$^4$\,L$_\odot$, which corresponds to an absolute magnitude of $\sim$\,--7.0; see Figure\,\ref{fig:detect}. This constraint on the outburst luminosity is within the luminosity range of RSGs \citep{Davies_2020_RSG}. Furthermore, this constraint corresponds to an apparent magnitude of $\sim$\,22, deeper than most of our upper limits. We additionally note that our model can be understood as a lower limit--if another power source contributed to the eruptions (e.g., recombination), we would expect brighter and longer-duration transients for a given set of parameters.

 Other investigations into pre-SN outbursts in SN\,2023ixf also have not found evidence for any detectable signatures \citep{Flinner_2023, 2023arXiv230813101P, Neustadt_2023_ixf}, although to varying limits. Our outburst constraints and photometric limits are comparable to those found by \citet{Dong_2023_ixf}, who derive an upper limit to the ejecta mass of 0.015\,M$_\odot$ based on the models of \citet{Tsuna_2023} (compared to our ejecta mass limit of $\textless$\,0.3\,M$_\odot$) for a hydrodynamical model that had peak $M_r\,\simeq\,-8$.
 
 When compared to SN\,2020tlf, any SN\,2023ixf pre-SN outburst would be fainter than the activity seen prior to SN\,2020tlf. On average, our limits are fainter than the pre-SN outburst of SN\,2020tlf by $\sim$\,2.5\,mag. 

 Defining the duration of a model outburst as the amount of time the outburst is brighter than detection limits, we find that the typical duration of a detectable outburst is similar, or shorter than the the gaps between the Pan-STARRS observations. The duration of an outburst at our upper luminosity and ejected mass limit is $\sim$\,100 days. This is shorter than the largest gap in the Pan-STARRS data of $\sim$\,600 days and there are multiple large gaps of over 100 days in the pre-explosion dataset. A detectable outburst may therefore not be detected due to larger gaps in the photometric coverage.
 
 In Figure\,\ref{fig:detect} we also show the luminosity that corresponds to the 80\% cutoff for bump detection and the corresponding luminosity of our averaged Pan-STARRS limits. Furthermore, we plot the upper values of the CSM mass for SN\,2023ixf \citep{Jacobson-Galan_2023_ixf} and SN\,2020tlf \citep{Jacobson-Galan_2020}. The upper value for the CSM mass from \citet{Jacobson-Galan_2023_ixf} which was derived from best-fit \texttt{CMFGEN} radiative transfer models is 0.07\,M$_\odot$, below our 80\% detection ejecta mass of 0.3\,M$_\odot$. Our limit is consistent with the CSM mass estimated by \citet{Kilpatrick_2023_ixf} who found a dusty CSM mass of $\sim$\,5\,$\times$\,10$^{-5}$\,M$_\odot$ and \citet{SinghTeja_2023} find a CSM mass between 0.001 and 0.030\,M$_\odot$. Similarly, \citet{2023arXiv230813101P} constrain the mass loss rate of the progenitor from their X-ray analysis to $\lesssim$\,5\,$\times$\,10$^{-4}$\,M$_\odot$\,yr$^{-1}$, consistent with our limit. \citet{Hiramatsu_2023} estimate mass loss rates of 0.1\,--\,1.0\,M$_\odot$\,yr$^{-1}$ in the 1-2 years before the SN explosion using numerical light curve models informed by early followup observations.  \citet{Qin_2023} used the archival \textit{HST} and \textit{Spitzer} imaging to examine the progenitor of SN\,2023ixf, finding a mass loss rate of $\sim$\,3.6\,$\times$\,10$^{-4}$\,M$_\odot$\,yr$^{-1}$, concluding that this enhanced mass loss rate (compared to RSG winds) was consistent with there being pulsational mass loss. \citet{Jencson_2023_ixf} also conclude enhanced mass loss rates deduced from their IR analysis of the progenitor of SN\,2023ixf, finding that the mass loss rate of the progenitor 3\,--\,19 years prior to explosion was $\sim$\,3\,$\times$\,10$^{-5}$\,--\,3\,$\times$\,10$^{-4}$\,M$_\odot$\,yr$^{-1}$. Using a period-luminosity relation with the IR variability of the progenitor of SN\,2023ixf, \citet{Soraisam_2023_ixf} found a mass loss rate of 2\,--\,4\,$\times$\,10$^{-4}$\,\,M$_\odot$\,yr$^{-1}$. In short, all mass-loss rate estimates seem consistent with our limit of 0.3\,M$_\odot$ of ejected mass in an outburst forming the CSM.

We repeat our eruption-search methodology utilizing the radiation hydrodynamic models of pre-SN outbursts in SNe\,II devised by \citet{Tsuna_2023}. We select the two extreme models in terms of luminosity: the ``double-large", corresponding to 3.6\,M$_\odot$ of CSM and 1.4\,$\times$\,10$^{47}$\,erg in radiated energy; and the ``single-small" model, being the least energetic and corresponding to an ejected mass of 0.015\,M$_\odot$ and radiated energy of 2.0\,$\times$\,10$^{45}$\,erg. When using these models to construct training set light curves, our only free parameter is the time of explosion. Again, we create a training set of 10$^4$ model light curves and add appropriate extinction to these light curves \citep[which was not considered in the initial modelling by ][]{Tsuna_2023}. The resultant classifier was then applied to our long baseline pre-explosion data. Our classifier, again, does not detect pre-SN eruptions consistent with this model. This is consistent with the analysis of \citet{Dong_2023_ixf}, who do not find any of the models of \citet{Tsuna_2023} to likely be represented in their pre-explosion data. The top row of Figure\,\ref{fig:detect} also shows the single-small and double-long models (the least and most luminous of their hydrodynamical pre-explosion outburst models) of \citet{Tsuna_2023} for reference. With a peak at $\sim$\,--10.5 and duration of a few hundred days in the case of the double-long model, our Pan-STARRS observations would be sensitive to outbursts that follow this model.

\subsection{Pre-explosion variability of the progenitor} \label{sec:vary}

Numerous previous studies of the pre-explosion activity of SN\,2023ixf found that the progenitor was observably variable in the IR wavelengths \citep[see;][]{Kilpatrick_2023_ixf, Soraisam_2023_ixf, Jencson_2023_ixf}. \citet{Kilpatrick_2023_ixf} suggested that the variability, with a period of around 1,000 days seen in the pre-explosion \textit{Spitzer} data may be due to the $\kappa$-mechanism pulsations seen in RSGs such as $\alpha$\,Ori \citep[Betelgeuse, see;][]{Li_1994, Heger_1997}, where changing opacity drives variability. Apart from deep \textit{HST} single epoch images, in optical bands, the progenitor is not detected. However, we may extend our methodology to place constraints on the variability of the progenitor in the optical. 

Similarly to our pre-SN outburst model, we construct a simple variability model, assuming sinusoidal variability, anti-phase to the IR variability (i.e. assuming constant bolometric luminosity). This model has a fixed period of 1000 days and two free parameters, the amplitude of the variation and the baseline. Again, we train an multilayer perceptron with the same number of layers, number of neurons and the same activation function as in Section\,\ref{sec:net}. We randomly sample both the amplitude and baseline between 0 and 10$^6$\,L$_\odot$ and create 10$^4$ test lightcurves (both with and without variability) with which we construct our training set. We then run the $i$-band Pan-STARRS pre-explosion data through this model. We choose the $i$-band as this has the most data and best temporal coverage, also using one filter avoids making assumptions on color-evolution. In the pre-explosion data, we find \textit{no} detectable variability in the $i$-band data.

To place upper limits on the variability, we repeat the methodology used to constrain the pre-SN outbursts (see Figure\,\ref{fig:detect}). We vary the baseline and amplitude (separately) between 0\,--\,10$^6$\,L$_\odot$ with each step having 10$^3$ test LCs generated. For each set of 10$^3$ LCs, the other unfixed parameter is varied randomly. Using the same 80\% detection efficiency threshold, we find that these models are not sensitive to the baseline and the amplitude has an upper limit of $\sim$\,4\,$\times$\,10$^4$\,L$_\odot$. This limit is similar to the constraint from the pre-SN outburst models and is similar to the luminosity of RSG progenitors. This suggests that if our optical images were close to the depth of the progenitor, we would have observed variability.

Moreover, we vary our SED models to infer the limits of variability in other bands (in a non-periodic fashion). We use the RSG progenitor parameters derived from our SED analysis using the consolidated photometry presented in section\,\ref{sec:prog}. Firstly, we vary only the dust properties of the progenitor with the other parameters being fixed. We vary the optical depth, $\tau$, between 2\,--\,10. We then test a second scenario in which the progenitor properties (luminosity and temperature) are freely varied, with a fixed $\tau$\,=\,8.23 (the value from our SED fitting). In these two tests, we use the \textit{Spitzer} observations to constrain the remaining free parameters; we use a Gaussian Process interpolation to predict the \textit{Spitzer} observed fluxes throughout the observed baseline. 

\begin{figure}[!t]
	\includegraphics[width=0.99\columnwidth]{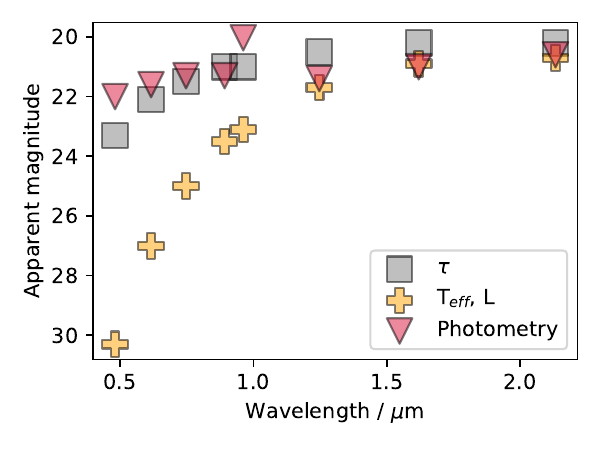}
    \caption{The peaks of the possible pre-explosion variability in the Pan-STARRS \textit{grizy} and also the near-IR \textit{JHK} filters (red triangles). The two methods used to fit the variability are compared with the photometric limits. Grey squares show the optical depth of the CSM as the driver for variability, while yellow crosses show the RSG properties as the cause.}
    \label{fig:vary}
\end{figure}

\begin{figure}[!t]
	\includegraphics[width=0.99\columnwidth]{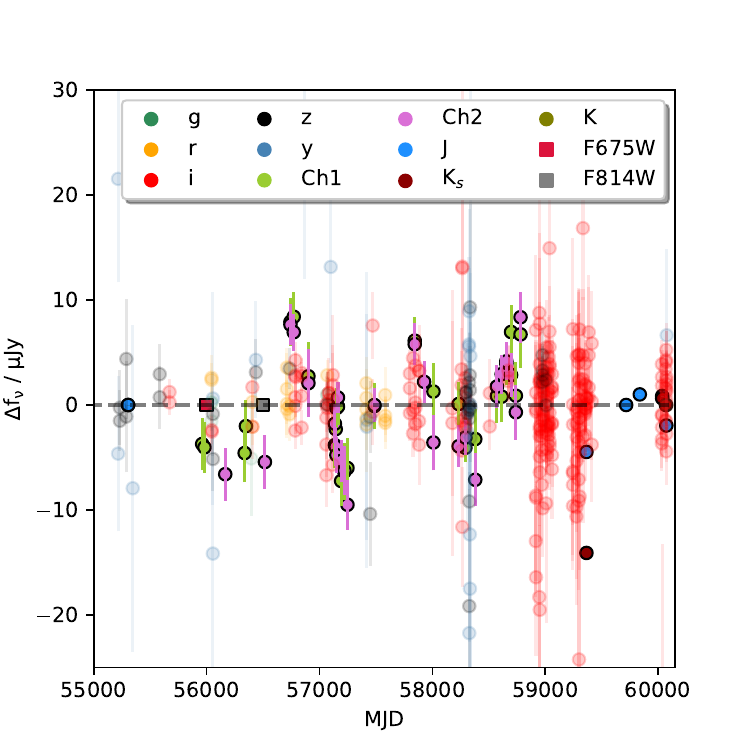}
    \caption{The scatter of the photometry in the pre-explosion light curve of the progenitor of SN\,2023ixf. Plotted is the scatter in the flux measurements of the Pan-STARRS data, \textit{HST} detections, \textit{Spitzer} and near-IR data. The \textit{HST} observations are placed at arbitrary dates. The dashed horizontal line represents a $\Delta$f$_\mathrm{\nu}$\,=\,0.}
    \label{fig:scatter}
\end{figure}

The peaks of the variability for each \textit{grizy} filter with each method and the limits from our photometry are shown in Figure\,\ref{fig:vary}. When the variability is accounted for by changing the progenitor parameters, the variability never peaks brighter than our limits. When the variability is assumed to be due to changes in the optical depth, in the optical, all but the $z$-band have photometric limits brighter than the peak of the variability. This may suggest that our $z$-band photometric coverage did not catch a peak in the variability if it was detectable or that the variability may not be purely due to optical depth variations. Generally, we would not have been able to detect variability of the progenitor of SN\,2023ixf in the framework of our assumptions with Pan-STARRS. Also shown in Figure\,\ref{fig:vary} are the near-IR bands, \textit{JHK}. The progenitor of SN\,2023ixf was detected in the near-IR; however, these detections occur \textit{outside} of the \textit{Spitzer} baseline. Nevertheless, these observations are similar to the peaks of the variability in both scenarios, being dimmer than the peak of the variability when just the optical depth is varied and brighter than the case where the progenitor properties are free parameters. For variability in the IR, we also note that the fractional variability, defined as the range in flux measurements over the baseline (taken as the average flux) is approximately constant over all IR filters. The scatter in the flux measurements is presented in Fig.\,\ref{fig:scatter}. Systematically adding to the uncertainty of each measurement in quadrature (adding fractional uncertainty of 0.0001 each step) to represent intrinsic scatter allows us to probe possible variability. By calculating how much scatter is required to produced a reduced $\chi^2$\,=\,1, compared with zero scatter, $\Delta$f$_\mathrm{\nu}$\,=\,0\,$\mu$Jy, we can estimate the intrinsic scatter. In the Pan-STARRS \textit{izy} filters (the filters with the most flux measurements), typically $\lesssim$\,5$\%$ of the uncertainty is required to be added as intrinsic scatter. This may indicate some marginal variability in these data. However, we note that there may be underestimates in the uncertainties in this analysis and that the typical uncertainty of the flux measurements is larger than the the typical IR variability.


\section{The host, M\,101, the Pinwheel Galaxy} \label{sec:host}

\begin{figure*}[t]
	\includegraphics[width=0.99\textwidth]{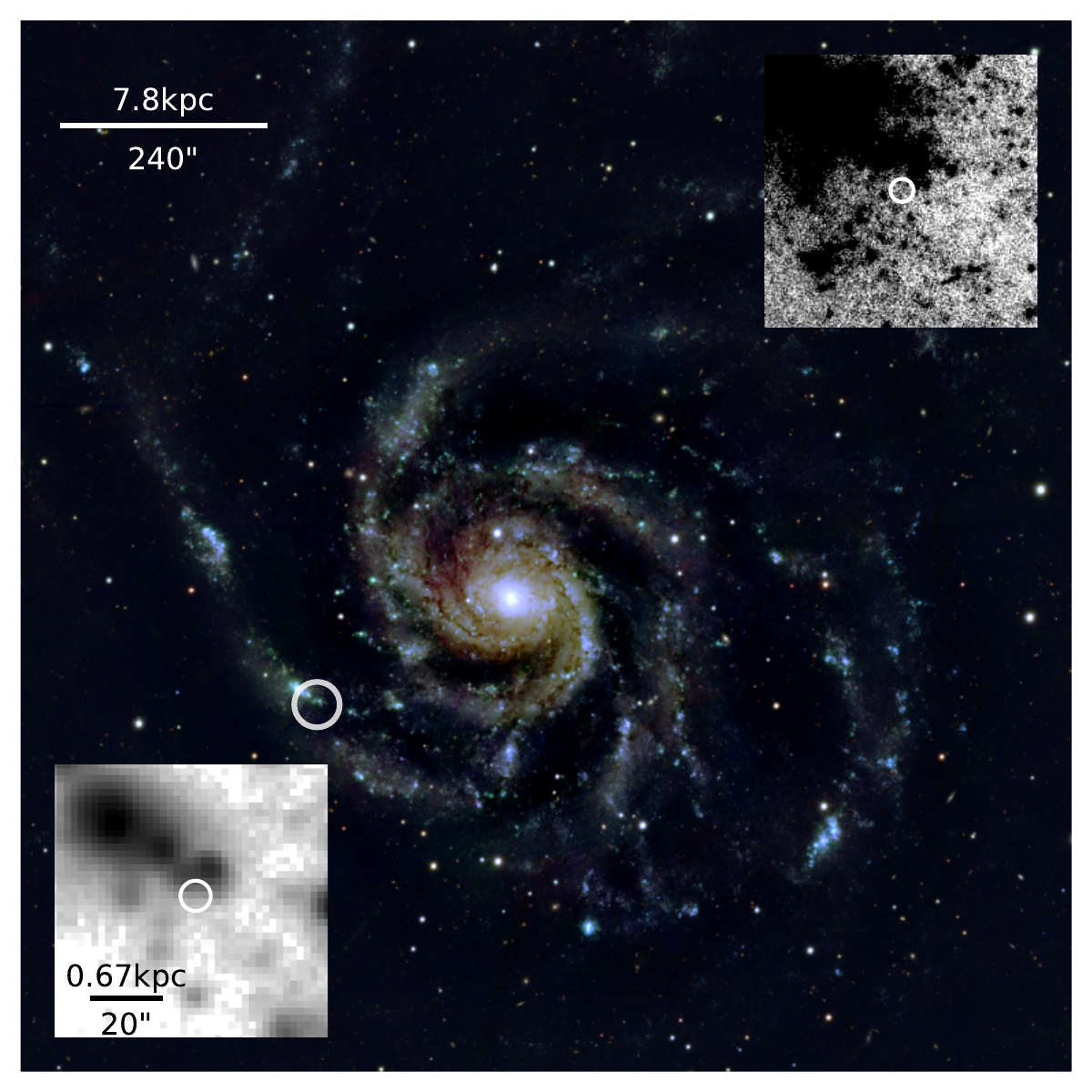}
    \caption{The host of SN\,2023ixf, M\,101 or NGC\,5457. This is a color-composite image made with PS1 $gri$ images. In both the full host image and the inset ``zoomed-in" images, the location of SN\,2023ixf is denoted by a white circle. The host image was created using mosaiced $gri$ images from the Pan-STARRS image cutout service\footnote{http://ps1images.stsci.edu/cgi-bin/ps1cutouts}. The inset image in the upper-right corner is an $i$-band stacked image from  the Pan-STARRS image cutout service which has the same pixel scale as the color image. The cutout on the bottom left is the location of SN\,2023ixf in a continuum subtracted H$\alpha$ image of M\,101 from \citet{Hoopes_2001}, downloaded from the NASA/IPAC Extragalactic Database (NED). All images are aligned North as up and East to the left.}
    \label{fig:host}
\end{figure*}

 The host, Messier 101 (M\,101), also known as NGC\,5457, or the Pinwheel Galaxy is located at a redshift of 0.000804 \citep{Perley_2023ixf} and is a face-on spiral galaxy \citep[SABc;][]{m101_class}. As can be seen in Figure\,\ref{fig:host}, SN\,2023ixf is coincident with a spiral arm at an offset of 264" ($\sim$\,8.7\,kpc) from the center of the nucleus of the host. SN\,2023ixf is the fifth recorded SN in M\,101 the others being: SN\,2011fe \citep{Nugent_2011}; SN\,1970G \citep{1970g}; SN\,1951H \citep[e.g.][]{1951h} and SN\,1909A \citep{1909a}. 

In order to gauge the association of the location of SN\,2023ixf with the local star-formation, we utilize the pixel statistics technique, that takes advantage of a normalized cumulative ranking \citep[NCR, see;][for details on this method]{ja06, Ransome_2022}. This technique has been used to compare the environments of different SN classes with star formation as traced by H$\alpha$ emission \citep{ja06, and12, hab14, Ransome_2022}.  In short, NCR processing consists of sorting a continuum subtracted image by pixel (flux) value, cumulatively summed and normalized by the total (e.g., each pixel now has a value between 0 and 1).

We show the ``NCR image" of the local environment of as the bottom left inset of Figure\,\ref{fig:host}. This continuum subtracted H$\alpha$ image was downloaded from the NASA/IPAC Extragalactic Database (NED\footnote{http://ned.ipac.caltech.edu}), with the original observations by \citet{Hoopes_2001} at the Kitt Peak National Observatory Burrel Schmidt Telescope. After NCR processing, we find that the NCR value at the site of SN\,2023ixf is 0.27\,$\pm$\,0.08. This NCR value is almost identical to the average NCR value of SNe\,IIP presented by \citet{and12} of 0.26, who measured the NCR value from observations of the hosts of 58 SNe\,IIP. Therefore the environment of SN\,2023ixf in terms of association to star formation traced by H$\alpha$ is unremarkable for SNe\,II.

\section{Conclusions and summary} \label{sec:conc}

In this work, we have presented the long-baseline pre-explosion light curve of the nearby SN\,2023ixf in M\,101 as observed by Pan-STARRS. With limits from this photometry and stacked images and also measurements from the literature, we find a progenitor consistent with a RSG with mass 14\,--\,20\,M$_\odot$, in agreement with most previous work. Using neural net classifiers, we do not find evidence of outbursts that may have produced the confined CSM but were able to place limits on any possible outbursts. Our findings can be summarized as follows:

\begin{itemize}

    \item Using our source injection photometric methodology for obtaining pre-explosion limits, we do not detect any pre-explosion activity in the Pan-STARRS \textit{grizy} filters. The average limits limits obtained are $M_g\,=\,-7.2$\,mag; $\,M_r=\,-7.8$\,mag; $M_i\,=\,-7.9$\,mag; $M_z\,=\,-7.9$\,mag and $M_y\,=\,-9.2$\,mag. These limits are below the brightness of the pre-SN outburst seen in SN\,2020tlf and much fainter than outbursts seen prior to SNe\,IIn with pre-explosion outburst detections. Therefore, if the progenitor of SN\,2023ixf suffered an outburst similar to previous observed events (with a duration of $\>100$ days), Pan-STARRS would have been able to detect it, if the outburst didn't occur during a gap in the data.

    \item We train a multilayer perceptron using an expanding photosphere model and the model outlined in \citet{Tsuna_2023} to identify outbursts in our Pan-STARRS light curves. We do not find evidence for these types of outbursts in our pre-SN data.
    
    \item Using our multilayer perceptron classifier, we find that our outburst luminosity has an upper limit absolute magnitude of $\sim$\,--7.0 and ejecta mass to less than 0.3\,M$_\odot$. These constraints are consistent with measurements from the literature.
    
    \item Multi-year deep stacks in the $wizy$ bands do not yield a progenitor detection to 3$\sigma$ limits of $w$\,=\,24.80\,mag, $i$\,=\,23.80\,mag, $z$\,=\,23.00\,mag and $y$\,=\,20.03\,mag. This is consistent our best-fit progenitor SED and shallower than the optical \textit{HST} detections.

    \item We train another multilayer perceptron to detect periodic variability, following the period discovered in \textit{Spitzer} observations. We do not detect any pre-SN variability in the most sampled filter, $i$ using the neural network. We repeat the methodology used for the pre-SN outburst to place limits on variability, finding similar limits on the amplitude of the variation as we found with the pre-SN outburst model.
    
    \item We fit SEDs using \texttt{DUSTY+}MARCS models to consolidated literature photometry of the progenitor with conservative uncertainty estimates to account its variability in IR wavelengths. We use a carbon dust model and find a progenitor mass range of 14\,--\,20\,M$_\odot$. This mass range is consistent with other reported values for SN\,2023ixf from the literature and may indicate a RSG progenitor on the higher end of the observed mass range.

    \item By varying both the dust properties and progenitor temperature and luminosity and fitting to the SEDs of a varying progenitor, we find that optical variability consistent with \textit{Spitzer} observations and our \texttt{DUSTY} models was not observable with Pan-STARRS. 
    
    \item Using a NCR pixel statistics method, we find that the host environment of SN\,2023ixf, with an NCR value of 0.27\,$\pm$\,0.08, is consistent with the average NCR value of the environments of SN\,IIP and indicative of an environment of moderate ongoing star-formation. 
\end{itemize}



\noindent \textbf{ACKNOWLEDGMENTS}\\
\\
We thank Daichi Tsuna for valuable discussion and providing their pre-SN outburst models. We also thank Sarah McDonald for helpful discussions.
 
V.A.V.\ and C.L.R.\ acknowledge support from the Charles E.\ Kaufman Foundation through the New Investigator Grant KA2022-129525.
W.J.-G.\ is supported by the National Science Foundation Graduate Research Fellowship Program under Grant No.~DGE-1842165.
C.D.K.\ acknowledges partial support from a CIERA postdoctoral fellowship.
M.G.\ acknowledges support from the European Union’s Horizon 2020 research and innovation programme under ERC Grant Agreement No.\ 101002652 and Marie Sk\l{}odowska-Curie Grant Agreement No. 873089.
M.R.D.\ acknowledges support from the NSERC through grant RGPIN-2019-06186, the Canada Research Chairs Program, and the Dunlap Institute at the University of Toronto.
C.G.\ is supported by a VILLUM FONDEN Young Investigator Grant (project number 25501).
R.Y.\ is grateful for support from a Doctoral Fellowship from the University of California Institute for Mexico and the United States (UCMEXUS) and a NASA FINESST award (21-ASTRO21-0068).  The UCSC team is supported in part by NASA grant NNG17PX03C, NSF grant AST--1815935, the Gordon \& Betty Moore Foundation, the Heising-Simons Foundation, and by a fellowship from the David and Lucile Packard Foundation to R.J.F.

The Young Supernova Experiment (YSE) and its research infrastructure is supported by the European Research Council under the European Union's Horizon 2020 research and innovation programme (ERC Grant Agreement 101002652, PI K.\ Mandel), the Heising-Simons Foundation (2018-0913, PI R.\ Foley; 2018-0911, PI R.\ Margutti), NASA (NNG17PX03C, PI R.\ Foley), NSF (AST-1720756, AST-1815935, PI R.\ Foley; AST-1909796, AST-1944985, PI R.\ Margutti), the David \& Lucille Packard Foundation (PI R.\ Foley), VILLUM FONDEN (project 16599, PI J.\ Hjorth), and the Center for AstroPhysical Surveys (CAPS) at the National Center for Supercomputing Applications (NCSA) and the University of Illinois Urbana-Champaign.

Pan-STARRS is a project of the Institute for Astronomy of the University of Hawaii, and is supported by the NASA SSO Near Earth Observation Program under grants 80NSSC18K0971, NNX14AM74G, NNX12AR65G, NNX13AQ47G, NNX08AR22G, 80NSSC21K1572 and by the State of Hawaii.  The Pan-STARRS1 Surveys (PS1) and the PS1 public science archive have been made possible through contributions by the Institute for Astronomy, the University of Hawaii, the Pan-STARRS Project Office, the Max-Planck Society and its participating institutes, the Max Planck Institute for Astronomy, Heidelberg and the Max Planck Institute for Extraterrestrial Physics, Garching, The Johns Hopkins University, Durham University, the University of Edinburgh, the Queen's University Belfast, the Harvard-Smithsonian Center for Astrophysics, the Las Cumbres Observatory Global Telescope Network Incorporated, the National Central University of Taiwan, STScI, NASA under grant NNX08AR22G issued through the Planetary Science Division of the NASA Science Mission Directorate, NSF grant AST-1238877, the University of Maryland, Eotvos Lorand University (ELTE), the Los Alamos National Laboratory, and the Gordon and Betty Moore Foundation.

%

\vspace{5mm}
\facilities{Pan-STARRS}


\software{astropy \citep{astropy}, numpy \citep{numpy}, Photpipe \citep{Rest+05}, tensorflow, \citep{tensorflow}, pandas \citep{pandas}  
          }





\bibliography{sample631}{}
\bibliographystyle{aasjournal}



\end{document}